\documentclass[11pt]{iopart}
\usepackage[utf8]{inputenc}
\usepackage{multicol}
\usepackage[T1]{fontenc}
\expandafter\let\csname equation*\endcsname=\relax
\expandafter\let\csname endequation*\endcsname=\relax
\usepackage{amsmath,amssymb,amsfonts,amsthm}

\catcode`,\active

\catcode`\,12

\newcommand*\pFqskip{8mu}
\catcode`,\active
\newcommand*\pFq{\begingroup
        \catcode`\,\active
        \def ,{\mskip\pFqskip\relax}%
        \dopFq
}
\catcode`\,12
\def\dopFq#1#2#3#4#5{%
        {}_{#1}F_{#2}\biggl(\genfrac..{0pt}{}{#3}{#4};#5\biggr)%
        \endgroup
}
\newcommand{\Keywords}[1]{\par\noindent
{\footnotesize{\textbf{Keywords}\/}: #1}}
\newcommand{\nombrespacs}[1]{\par\noindent
{\small{PACS numbers\/}: #1}}
\newcommand{\class}[1]{\par\noindent
{\small{AMS classification scheme numbers\/}: #1}}

\newcommand{\ket}[1]{|#1\rangle}
\newcommand{\kket}[2]{|#1\rangle_{#2}}
\newcommand{\bra}[1]{\langle #1|}

\newcommand{\Braket}[3]{\bra{#1}#2\ket{#3}}
\newcommand{\BBraket}[5]{{}_{#1}\bra{#2}#3\ket{#4}_{#5}}
\newcommand{\braket}[2]{\langle #1|#2\rangle}
\newcommand{\bbraket}[4]{{}_{#1}\langle #2|#3\rangle_{#4}}
\numberwithin{equation}{section}
\begin{document}
\title[Interbasis expansion and bivariate Krawtchouk polynomials]{Interbasis expansions for the isotropic 3D harmonic oscillator and bivariate Krawtchouk polynomials}
\author{Vincent X. Genest}
\ead{genestvi@crm.umontreal.ca}
\address{Centre de recherches math\'ematiques, Universit\'e de Montr\'eal, C.P. 6128, Succursale Centre-ville, Montr\'eal, Qu\'ebec, Canada, H3C 3J7}
\author{Luc Vinet}
\ead{luc.vinet@umontreal.ca}
\address{Centre de recherches math\'ematiques, Universit\'e de Montr\'eal, C.P. 6128, Succursale Centre-ville, Montr\'eal, Qu\'ebec, Canada, H3C 3J7}
\author{Alexei Zhedanov}
\ead{zhedanov@yahoo.com}
\address{Donetsk Institute for Physics and Technology, Donetsk 83114, Ukraine}
\begin{abstract}
An explicit expression for the general bivariate Krawtchouk polynomials is obtained in terms of the standard Krawtchouk and dual Hahn polynomials. The bivariate Krawtchouk polynomials occur as matrix elements of the unitary reducible representations of $SO(3)$ on the energy eigenspaces of the 3-dimensional isotropic harmonic oscillator and the explicit formula is obtained from the decomposition of these representations into their irreducible components. The decomposition entails expanding the Cartesian basis states in the spherical bases that span irreducible $SO(3)$ representations. The overlap coefficients are obtained from the Clebsch-Gordan problem for the $\mathfrak{su}(1,1)$ Lie algebra.\\

\\
\vfill
\nombrespacs{03.65.Fd, 02.20.-a}\\
\class{33C50, 33C80, 81Q80}
\end{abstract}

\section{Introduction}
The standard Krawtchouk polynomials orthogonal with respect to the binomial distribution are known to enter the expression of the Wigner $\mathcal{D}$-functions which give the matrix elements of the irreducible representations of $SU(2)$ in the standard bases. The multivariate polynomials that generalize them are orthogonal with respect to the multinomial distribution. Although the definition of the multivariate polynomials goes back to 1971 when it was given in a Statistics context \cite{Griffiths-1971}, to our knowledge their introduction in the study of Mathematical Physics problems is much more recent (see \cite{Genest-2013-1} for more background). For instance, the bivariate Krawtchouk polynomials have been seen to occur in the wavefunctions of a superintegrable finite oscillator model with $SU(2)$ symmetry \cite{Miki-2013}. They have also been shown to arise as the $9j$-symbol of the oscillator algebra \cite{Zhedanov-1997}. As well, the 2-variable Krawtchouk polynomials have been used to design a two-dimensional spin lattice with remarkable quantum state transfer properties \cite{Miki-2012}.

Lately, the Krawtchouk polynomials in $n$ discrete variables have been interpreted as matrix elements of the reducible representations of $SO(n+1)$ on the energy eigenspaces of the $(n+1)$-dimensional isotropic harmonic oscillator \cite{Genest-2013-1}. This has provided a natural setting within which the various properties of these polynomials could be straightforwardly derived. It is the purpose of this paper to further exploit this group theoretical connection and to obtain a new expansion formula that emerges from the irreducible decomposition of the relevant rotation group representations. The overlap coefficients between the Cartesian and spherical bases \cite{Pogo-2006} will be needed and it shall also be indicated how these can be recovered using a correspondence with the Clebsch-Gordan problem of the $\mathfrak{su}(1,1)$ algebra. The focus here is on the bivariate case.

\subsection{Three-dimensional isotropic harmonic oscillator}
The isotropic 3-dimensional harmonic oscillator is described by the Hamiltonian
\begin{align}
\label{3D-Hamiltonian}
\mathcal{H}=-\frac{1}{2}\nabla^2+\frac{1}{2}(x^2+y^2+z^2),
\end{align}
where $\nabla^2$ denotes the Laplacian. The Schr\"odinger equation $\mathcal{H}\Psi=\mathcal{E}\Psi$ associated to \eqref{3D-Hamiltonian} separates in particular in Cartesian, polar (cylindrical) and spherical coordinates. In each of these coordinate systems, the exact solutions are known \cite{Cohen-1991} and the eigenstates of \eqref{3D-Hamiltonian} are labeled by three quantum numbers. One has the following bases and the corresponding wavefunctions for the states of the oscillator:
\begin{enumerate}
\item The Cartesian basis denoted by $\ket{n_x,n_y,n_z}_{C}$ where $n_x,n_y,n_z\in \mathbb{N}$ and with energy eigenvalue $\mathcal{E}=n_x+n_y+n_y+3/2=N+3/2$. The associated wavefunctions are denoted $\Psi_{n_x,n_y,n_z}(x,y,z)$ and given by
\begin{align}
\label{Cartesian-Wave}
\begin{aligned}
&\Psi_{n_x,n_y,n_z}(x,y,z)=
\\
&\sqrt{\frac{1}{2^{N}\pi^{3/2}n_x!n_y!n_{z}!}}\,e^{-(x^2+y^2+z^2)/2}\,H_{n_x}(x)\,H_{n_y}(y)\,H_{n_z}(z),
\end{aligned}
\end{align}
where $H_{n}(x)$ stands for the Hermite polynomials \cite{Koekoek-2010}.
\item The polar basis denoted by $\ket{n_{\rho},m,n_z}_{P}$ where $n_{\rho}\in \mathbb{N}$, $m\in \mathbb{Z}$, $n_z\in \mathbb{N}$ and with energy eigenvalue $\mathcal{E}=2n_{\rho}+|m|+n_z+3/2=N+3/2$. The associated wavefunctions are denoted $\Psi_{n_{\rho},m,n_{z}}(\rho,\phi,z)$ and given by
\begin{align}
\label{Polar-Wave}
\begin{aligned}
&\Psi_{n_{\rho},m,n_z}(\rho,\phi,z)=
\\
&
\frac{(-1)^{n_{\rho}}}{\pi^{3/4}}\sqrt{\frac{n_{\rho}!}{2^{n_z}n_z!\,\Gamma(n_{\rho}+|m|+1)}}\,e^{-(\rho^2+z^2)/2}\rho^{|m|}L_{n_{\rho}}^{(|m|)}(\rho^2)H_{n_z}(z) e^{im\phi},
\end{aligned}
\end{align}
where  $L_{n}^{(\alpha)}(x)$ are the Laguerre polynomials \cite{Koekoek-2010}.
\item The spherical basis $\ket{n_r,\ell,m}_{S}$ where $n_r\in\mathbb{N}$, $\ell\in\mathbb{N}$,  $m=-\ell,\ldots,\ell$ and with energy eigenvalue $\mathcal{E}=2n_r+\ell+3/2=N+3/2$. The wavefunctions are denoted $\Psi_{n_r,\ell,m}(r,\theta,\phi)$ and given by
\begin{align}
\label{Spherical-Wave}
\begin{aligned}
&\Psi_{n_r,\ell,m}(r,\theta,\phi)=
\\
&(-1)^{n_r}\,e^{-r^2/2}\,r^{\ell}\,\sqrt{\frac{2n_{r}!}{\Gamma(n_r+\ell+3/2)}}\,L_{n_r}^{(\ell+1/2)}(r^2)\,Y_{\ell}^{m}(\theta,\phi),
\end{aligned}
\end{align}
where $Y_{\ell}^{m}(\theta,\phi)$ are the spherical harmonics \cite{Cohen-1991}.
\end{enumerate}
It is directly seen that the energy level $N$ has degeneracy $(N+1)(N+2)/2$.
The creation/annihilation operators
\begin{align*}
a_{x_i}=\frac{1}{\sqrt{2}}(x_i+\partial_{x_i}),\quad a_{x_{i}}^{\dagger}=\frac{1}{\sqrt{2}}(x_i-\partial_{x_i}),\quad i=1,2,3,
\end{align*}
with $x_1=x$, $x_2=y$, $x_3=z$ obey the commutation relations
\begin{align*}
[a_{x_i},a_{x_{j}}^{\dagger}]=\delta_{ij},\quad [a_{x_i},a_{x_j}]=0,\qquad i,j=1,2,3,
\end{align*}
and have the following actions on the Cartesian basis states:
\begin{align*}
a_{x_{i}}\ket{n_{x_i}}_{C}=\sqrt{n_{x_i}}\,\ket{n_{x_i}-1}_{C},\, a_{x_{i}}^{\dagger}\ket{n_{x_i}}=\sqrt{n_{x_i}+1}\,\ket{n_{x_i}+1}_{C}\, .
\end{align*}
It follows that $a_{x_{i}}^{\dagger}a_{x_i}\ket{n_{x_i}}=n_{x_i}\ket{n_{x_i}}_{C}$.
In terms of these operators, \eqref{3D-Hamiltonian} takes the form
\begin{align}
\label{3D-Hamiltonian-Op}
\mathcal{H}=a_{x}^{\dagger}a_{x}+a_{y}^{\dagger}a_{y}+a_{z}^{\dagger}a_{z}+3/2,
\end{align}
and one has indeed $\mathcal{H}\ket{n_x,n_y,n_z}_{C}=(N+3/2)\ket{n_x,n_y,n_z}_{C}$.
\subsection{$SO(3)\subset SU(3)$ and oscillator states}
The Hamiltonian \eqref{3D-Hamiltonian-Op} of the 3-dimensional isotropic Harmonic oscillator is clearly invariant under $SU(3)$ transformations, which are generated by the constants of motion of the form $a_{i}^{\dagger}a_{j}$. For each value of $N$, the Cartesian basis states $\ket{n_x,n_y,n_z}_{C}$ support the completely symmetric irreducible representation of $SU(3)$. The Hamiltonian \eqref{3D-Hamiltonian} is also manifestly invariant under $SO(3)\subset SU(3)$ transformations. These rotations are generated by the three angular momentum generators
\begin{align}
\label{Angular}
L_{x}=-i(a_{y}^{\dagger}a_{z}-a_{z}^{\dagger}a_{y}),\; L_{y}=-i(a_{z}^{\dagger}a_{x}-a_{x}^{\dagger}a_{z}),\;
L_{z}=-i(a_{x}^{\dagger}a_{y}-a_{y}^{\dagger}a_{x}),
\end{align}
obeying the commutation relations
\begin{align*}
[L_x,L_{y}]=iL_{z},\quad [L_{y},L_{z}]=iL_{x},\quad [L_{z},L_{y}]=iL_{x}.
\end{align*}
The representation of $SO(3)$ on the oscillator states with a given energy is reducible. The irreducible content of this representation can be found by examining the states $\ket{n_r,\ell,m}_{S}$ of the spherical basis. These states are the common eigenstates of the $\mathfrak{so}(3)$ Casimir operator $\vec{L}^2=L_x^2+L_y^2+L_z^2$ and of $L_{z}$ with eigenvalues
\begin{align*}
\vec{L}^2\ket{n_r,\ell,m}_{S}=\ell(\ell+1)\ket{n_r,\ell,m}_{S},\quad L_{z}\ket{n_r,\ell,m}_{S}=m\ket{n_r,\ell,m}_{S}.
\end{align*}
For each value of $n_r$, these states provide a basis for the $(2\ell+1)$-dimensional irreducible representation of $SO(3)$. Since $N=2n_r+\ell$, it follows that for a given $N$ the $SO(3)$ representation on the eigenstates of the isotropic oscillator contains once, each and every $(2\ell+1)$-dimensional irreducible representation of $SO(3)$ with $\ell=N,N-2,\ldots, 1\text{ or } 0$, depending on the parity of $N$. One notes that in the polar basis $\ket{n_r,m,n_z}_{P}$, the following operators are diagonal:
\begin{align*}
L_{z}\ket{n_{\rho},m,n_z}_{P}=m\ket{n_{\rho},m,n_z}_{P},\quad a^{\dagger}_{z}a_{z}\ket{n_{\rho},m,n_z}_{P}=n_z\ket{n_{\rho},m,n_z}_{P}.
\end{align*}

\subsection{Unitary representations of $SO(3)$ and bivariate Krawtchouk polynomials}
Let $R\in SO(3)$ and consider the unitary representation provided by
\begin{align}
\label{Unirep}
U(R)=\exp \left(\sum_{i,j=1}^{3}B_{ij}a_{i}^{\dagger}a_{j}\right),
\end{align}
where $B^{\top}=-B$ and $R=e^{B}$. It has been shown in \cite{Genest-2013-1} that the matrix elements of this unitary operator in the Cartesian basis have the expression
\begin{align}
\nonumber
\BBraket{C}{i,k,l}{U(R)}{r,s,t}{C}=W_{i,k;N}\,P_{r,s}(i,k;N),
\end{align}
where $i+k+l=N=r+s+t$ and where
\begin{align}
\label{Weight}
W_{i,k;N}=\binom{N}{i,k}^{1/2}\;R_{33}^{N}
\left(\frac{R_{13}}{R_{33}}\right)^{i}
\left(\frac{R_{23}}{R_{33}}\right)^{k},
\end{align}
with $\binom{N}{i,k}$ denoting the trinomial coefficients
\begin{align*}
\binom{N}{i,k}=\frac{N!}{i!k!(N-i-k)!}.
\end{align*}
The $P_{r,s}(i,k;N)$ are the general bivariate Krawtchouk polynomials which have for parameters the entries $R_{ij}$ of the $3\times 3$ rotation matrix $R\in SO(3)$. The polynomials $P_{r,s}(i,k;N)$ enjoy many interesting properties. They are orthonormal with respect to the trinomial distribution
\begin{align*}
\sum_{i+k\leqslant N}W_{i,k;N}^2\,P_{r,s}(i,k;N)P_{r',s'}(i,k;N)=\delta_{rr'}\delta_{ss'}.
\end{align*}
and have for generating relation
\begin{align*}
\left(1+\frac{R_{11}}{R_{13}}u+\frac{R_{12}}{R_{13}}v\right)^{i}
&\left(1+\frac{R_{21}}{R_{23}}u+\frac{R_{22}}{R_{23}}v\right)^{k}
\left(1+\frac{R_{31}}{R_{33}}u+\frac{R_{32}}{R_{33}}v\right)^{N-i-k}\\
&=\sum_{r+s\leqslant N}\binom{N}{r,s}^{1/2}P_{r,s}(i,k;N)\,u^{r}v^{s}.
\end{align*}
The polynomials $P_{r,s}(i,k;N)$ have an explicit formula in terms of Gel'fand-Aomoto hypergeometric series
\small
\begin{align*}
&P_{r,s}(i,k;N)=\binom{N}{r,s}^{1/2}
\left(\frac{R_{31}}{R_{33}}\right)^{r}
\left(\frac{R_{32}}{R_{33}}\right)^{s}
\\
&\times\sum_{\alpha+\beta+\gamma+\delta\leqslant N}\frac{(-r)_{\alpha+\beta}(-s)_{\gamma+\delta}(-i)_{\alpha+\gamma}(-k)_{\beta+\delta}}{\alpha!\beta!\gamma!\delta!(-N)_{\alpha+\beta+\gamma+\delta}}(1-u_{11})^{\alpha}(1-u_{21})^{\beta}(1-u_{12})^{\gamma}(1-u_{22})^{\delta},
\end{align*}
\normalsize
where $(a)_{n}=(a)(a+1)\cdots(a+n-1)$ stands for the Pochammer symbol and where
\begin{align*}
u_{11}=\frac{R_{11}R_{33}}{R_{13}R_{31}},\,u_{12}=\frac{R_{12}R_{33}}{R_{13}R_{32}},\,u_{21}=\frac{R_{21}R_{33}}{R_{23}R_{31}},\,u_{22}=\frac{R_{22}R_{33}}{R_{23}R_{32}}.
\end{align*}
\normalsize
The polynomials $P_{r,s}(i,k;N)$ have the following integral representation involving the Hermite polynomials:
\begin{align*}
&P_{r,s}(i,k;N)=\frac{R_{13}^{-i}R_{23}^{-k}R_{33}^{-l}}{2^{N}\pi^{3/2}N!}\binom{N}{r,s}^{1/2}\\
&\times\int_{\mathbb{R}^{3}}e^{-(x_1^2+x_2^2+x_3^2)}
H_{r}(\widetilde{x}_1)
H_{s}(\widetilde{x}_2)
H_{t}(\widetilde{x}_3)
H_{i}(x_1)
H_{k}(x_2)
H_{l}(x_3)
\,dx_1dx_2dx_3.
\end{align*}
where $N=i+k+l=r+s+t$ and $(\widetilde{x}_1,\widetilde{x}_2,\widetilde{x}_3)^{\top}=R^{\top}(x_1,x_2,x_3)^{\top}$. They can also be expressed as a sum over products of three standard Krawtchouk polynomials \eqref{Krawtchouk}. 

In the case $R_{12}=0$, the general bivariate Krawtchouk polynomials $P_{r,s}(i,k;N)$ reduce to the bivariate Krawtchouk polynomials $K_2(m,n;i,k;\mathfrak{p}_1,\mathfrak{p}_2;N)$ introduced by Tratnik in \cite{Tratnik-1991} (see also \cite{Iliev-2010} for their bispectral properties). These polynomials have the explicit expression
\begin{align*}
K_2(m,n;i,k;\mathfrak{p}_1,\mathfrak{p}_2;N)=
\frac{(n-N)_{m}(i-N)_{n}}{(-N)_{m+n}}\;K_m(i;\mathfrak{p}_1;N-n)K_n(k;\frac{\mathfrak{p}_2}{1-\mathfrak{p}_1};N-i),
\end{align*}
where $K_n(x;p;N)$ stands for the standard Krawtchouk polynomials
\begin{align}
\label{Krawtchouk}
K_{n}(x;p;N)=\pFq{2}{1}{-n,-n}{-N}{\frac{1}{p}},
\end{align}
and where ${}_pF_{q}$ denotes the generalized hypergeometric function \cite{Koekoek-2010}. The condition $R_{12}=0$ is ensured if $R$ is taken to be a product of two successive clockwise rotations $R=R_{x}(\theta)R_{y}(\chi)$ around the $x$ and $y$ axes, respectively. This rotation is unitarily represented by $U(R)=e^{i\theta L_x}e^{i\chi L_y}$ and one has \cite{Genest-2013-1}
\begin{align}
\label{Tratnik-1}
\BBraket{C}{i,k,l}{e^{i\theta L_x}e^{i\chi L_y}}{r,s,t}{C}=R_{33}^{-N}\,W_{i,k;N}\widetilde{W}_{r,s;N}\,K_2(r,s;i,k;\mathfrak{p}_1,\mathfrak{p}_2;N),
\end{align}
where $\widetilde{W}_{m,n;N}$ is given by \eqref{Weight} with the parameters of the rotation matrix $R$ replaced by their transpose. One has again $r+s+t=N=i+k+l$ and furthermore $\mathfrak{p}_1=R_{13}^2$ and $\mathfrak{p}_2=R_{23}^2$. The polynomials of Tratnik thus depend only on two parameters, as opposed to three parameters for the general polynomials $P_{r,s}(i,k;N)$. The reader is referred to \cite{Genest-2013-1} for the group theoretical characterization of the polynomials $P_{r,s}(i,k;N)$ and references on the multivariate Krawtchouk polynomials.

\subsection{The main result}
The stage has now been set for the statement of the main formula of this paper. The most general rotation $R\in SO(3)$, which depends on three parameters, can be taken of the form
\small
\begin{align*}
R=
\begin{pmatrix}
c_{\alpha}c_{\beta}c_{\gamma}-s_{\alpha}s_{\gamma} & -s_{\alpha}c_{\beta}c_{\gamma}-c_{\alpha}s_{\gamma} & s_{\beta}c_{\gamma}\\
c_{\alpha}c_{\beta}s_{\gamma}+s_{\alpha}c_{\gamma}& c_{\alpha}c_{\gamma}-s_{\alpha}c_{\beta}s_{\gamma} & s_{\beta}s_{\gamma}\\
-c_{\alpha}s_{\beta} & s_{\alpha}s_{\beta} & c_{\beta}
\end{pmatrix},
\end{align*}
\normalsize
where $c_{\theta}=\cos \theta$ and $s_{\theta}=\sin \theta$. This rotation is unitarily represented by the operator
\begin{align*}
U(R)=e^{-i\gamma L_{z}}e^{-i\beta L_{y}}e^{-i\alpha L_{z}}.
\end{align*}
The parameters $\alpha$, $\beta$ and $\gamma$ thus correspond to the Euler angles. The decomposition of the $SO(3)$ representation on the energy eigenspaces of the isotropic 3D harmonic oscillator in irreducible components amounts to the expansion of the Cartesian basis states $\ket{n_x,n_y,n_z}_{C}$ in the spherical basis states $\ket{n_r,\ell,m}_{S}$:
\begin{align*}
&\BBraket{C}{i,k,l}{U(R)}{r,s,t}{C}\\
&=\sum_{n_r,\ell,m}\sum_{n_r',\ell',m'}\bbraket{C}{i,k,l}{n_r',\ell',m'}{S}\BBraket{S}{n_r',\ell',m'}{U(R)}{n_r,\ell,m}{S}\bbraket{S}{n_r,\ell,m}{r,s,t}{C},
\end{align*}
where $i+k+l=N=r+s+t$. The following expression for the bivariate Krawtchouk polynomials $P_{r,s}(i,k;N)$ stems from this decomposition:
\begin{align}
\label{Main-Formula}
\begin{aligned}
P_{r,s}&(i,k;N)=W_{i,k;N}^{-1}
\\
&\times \sum_{\substack{n_r,\ell\\2n_r+\ell=N }}\sum_{m,m'=-\ell}^{\ell}\mathcal{D}^{(\ell)}_{mm'}(R)\;\bbraket{C}{i,k,l}{n_r,\ell,m'}{S}\;\bbraket{S}{n_r,\ell,m}{r,s,t}{C}.
\end{aligned}
\end{align}
The matrix elements $\mathcal{D}^{(\ell)}_{m'm}(R)=\BBraket{S}{n_r',\ell',m'}{U(R)}{n_r,\ell,m}{S}$ of the $\mathfrak{so}(3)$ Wigner $\mathcal{D}$-matrix are given by \cite{Koornwinder-1982}
\begin{align*}
\nonumber
&\mathcal{D}^{(\ell)}_{m'm}=\delta_{n_rn_r'}\delta_{\ell\ell'}\;e^{-i(\gamma m'+\alpha m)}
\\
&\times (-1)^{m'+\ell}\sin^{2\ell}\big(\frac{\beta}{2}\big) \tan^{m+m'}\big(\frac{\beta}{2}\big)\left[\binom{2\ell}{m+\ell}\binom{2\ell}{m'+\ell}\right]^{1/2} K_{m+\ell}\big(m'+\ell;\,\sin^2\frac{\beta}{2};\,2\ell\big).
\end{align*}
The overlap coefficients between the Cartesian and spherical bases are obtained by using the intermediary decomposition over the polar basis states and read
\begin{align}
\label{Spherical/Cartesian}
\begin{aligned}
&\bbraket{S}{n_r,\ell,m}{r,s,t}{C}=
\\
&\sum_{n_{\rho}}\frac{(-1)^{\widetilde{n}_r+n_{\rho}}(-i)^{m+|m|}(-\sigma_m i)^{s}}{\sqrt{2}}\,\mathcal{C}^{\frac{1/2+q_r}{2},\frac{1/2+q_{s}}{2},\frac{1+|m|}{2}}_{\widetilde{r},\widetilde{s},n_{\rho}}\,\mathcal{C}_{n_{\rho},\widetilde{t},n_r}^{\frac{1+|m|}{2},\frac{1/2+q_{t}}{2},\frac{\ell+3/2}{2}},
\end{aligned}
\end{align}
where $2n_{\rho}+|m|=r+s$, $2n_r+\ell=r+s+t$ and $w=2\widetilde{w}+q_{w}$ with $w=r,s,t$ and $q_{w}=0,1$. In \eqref{Spherical/Cartesian}, the square root factor should be omitted for $m=0$ and $\sigma_m=1$ if $m\geqslant0$ and $-1$ otherwise. The coefficients $\mathcal{C}$ are given by
\begin{align*}
\nonumber
\mathcal{C}^{\nu_1,\nu_2,\nu_{12}}_{n_1,n_2,n_{12}}&=\delta_{\nu_{12},\nu_1+\nu_2+x}\left[\frac{(2\nu_1)_{n_1}(2\nu_2)_{n_2}(2\nu_1)_{x}}{n_1!n_2!n_{12}!x!(2\nu_2)_{x}(2\nu_1+2\nu_2+2x)_{n_{12}}(2\nu_1+2\nu_2+x-1)_{x}}\right]^{1/2}\\
&\,\times (x+n_{12})!\; R_{n_1}(\lambda(x);2\nu_1-1,2\nu_2-1;n_1+n_2),
\end{align*}
with $x=n_1+n_2-n_{12}$, where $R_{n}(\lambda(x);\gamma,\delta;N)$ are the dual Hahn polynomials \cite{Koekoek-2010} (see \eqref{Dual-Hahn}). One has $\bbraket{S}{n_r,\ell,m}{r,s,t}{C}=\bbraket{C}{r,s,t}{n_r,\ell,m}{S}^{*}$, where $x^{*}$ denotes the complex conjugate of $x$. Note that in \eqref{Main-Formula}, the dependence of the polynomials $P_{r,s}(i,k;N)$ on the parameters is all contained in the Wigner function. 

The main formula \eqref{Main-Formula} can also be used for the special case $R_{12}=0$ corresponding to the Tratnik polynomials. Indeed, since one has $e^{i\theta L_x}e^{i\chi L_y}=e^{-i\frac{\pi}{2} L_y}e^{i\theta L_z}e^{i\chi L_y}e^{i\frac{\pi}{2} L_y}$, it follows that
\small
\begin{align}
\label{Tratnik-2}
\BBraket{C}{i,k,l}{e^{i\theta L_x}e^{i\chi L_y}}{r,s,t}{C}=(-1)^{l+t}\;
\BBraket{C}{l,k,i}{e^{i\theta L_z}e^{i\chi L_y}}{t,s,r}{C},
\end{align}
\normalsize
where $i+k+l=N=r+s+t$. The LHS of \eqref{Tratnik-2} is given by \eqref{Tratnik-1} in terms of the Tratnik polynomials and the RHS of \eqref{Tratnik-2} is given by \eqref{Main-Formula} with the Euler angles values $\gamma=-\theta$, $\beta=-\chi$, $\alpha=0$. The following relations have been used to obtain \eqref{Tratnik-2}:
\begin{align*}
\BBraket{C}{a',b',c'}{e^{i\frac{\pi}{2}L_y}}{r,s,t}{C}=(-1)^{t}\delta_{a't}\delta_{b's},\quad \BBraket{C}{i,k,l}{e^{-i\frac{\pi}{2}L_y}}{a,b,c}{C}=(-1)^{l}\delta_{ic}\delta_{kb}.
\end{align*}
These relations are special cases of the formulas derived in \cite{Genest-2013-1} (see section 8). 
\subsection{Outline}
The remainder of the paper is organized in a straightforward manner. In section 2, the essentials of the $\mathfrak{su}(1,1)$ Lie algebra and its Clebsch-Gordan problems are reviewed. In section 3, the explicit expressions for the overlap coefficients between the Cartesian, polar and spherical bases are derived using their identification as Clebsch-Gordan coefficients of $\mathfrak{su}(1,1)$. A discussion of the generalization to $d$ variables is found in the conclusion.
\section{The $\mathfrak{su}(1,1)$ Lie algebra and the Clebsch-Gordan problem}
In this section, the essential results on the $\mathfrak{su}(1,1)$ Lie algebra that shall be needed are reviewed. In particular, the Clebsch-Gordan coefficients for the positive discrete series of irreducible representations are derived by a recurrence method. These coefficients are known (see for example \cite{Vilenkin-1991}) and are presented here to make the paper self-contained.
\subsection{The $\mathfrak{su}(1,1)$ algebra and its positive-discrete series of representations}
The $\mathfrak{su}(1,1)$ algebra has for generators $J_{0}$, $J_{\pm}$  which satisfy the commutation relations
\begin{align*}
[J_0,J_{\pm}]=\pm J_{\pm},\quad [J_{+},J_{-}]=-2J_{0}.
\end{align*}
The Casimir operator, which commutes with all generators, is given by
\begin{align}
\label{Casimir}
Q=J_{0}^2-J_{+}J_{-}-J_{0}.
\end{align}
The positive-discrete series of irreducible representations of $\mathfrak{su}(1,1)$ are labeled by a positive number $\nu>0$ and are infinite-dimensional. They can be defined by the following actions of the generators on a canonical basis $\ket{\nu,n}$, where $n\in \mathbb{N}$:
\begin{subequations}
\label{Actions-2}
\begin{align}
J_0\ket{\nu,n}&=(n+\nu)\ket{\nu,n},\\
J_{+}\ket{\nu,n}&=\sqrt{(n+1)(n+2\nu)} \ket{\nu,n+1},\\
J_{-}\ket{\nu,n}&=\sqrt{n(n+2\nu-1)}\ket{\nu,n-1}.
\end{align}
\end{subequations}
The $\mathfrak{su}(1,1)$-modules spanned by the basis vectors $\ket{\nu,n}$, $n\in \mathbb{N}$, with actions \eqref{Actions-2} will be denoted by $V^{(\nu)}$. As expected from Schur's lemma, the Casimir operator \eqref{Casimir} acts as a multiple of the identity on $V^{(\nu)}$:
\begin{align}
Q\ket{\nu,n}=\nu(\nu-1)\ket{\nu,n}.
\end{align}
\subsection{The Clebsch-Gordan problem}
The vector space $V^{(\nu_1)}\otimes V^{(\nu_2)}$ is a module for the $\mathfrak{su}(1,1)$ algebra generated by
\begin{align}
\label{Addition-Rule}
J_0^{(12)}=J_{0}^{(1)}+J_{0}^{(2)},\quad J_{\pm}^{(12)}=J_{\pm}^{(1)}+J_{\pm}^{(2)},
\end{align}
where the superscripts indicate on which vector space the generators act, for example $J_{\pm}^{(2)}=1\otimes J_{\pm}$. In general, this module is not irreducible. From the addition rule \eqref{Addition-Rule}, it is easy to see that each irreducible representation occurs only once and hence that one has the irreducible decomposition
\begin{align}
\label{Decompo}
V^{(\nu_1)}\otimes V^{(\nu_2)}=\bigoplus_{\nu_{12}} V^{(\nu_{12})}.
\end{align}
The admissible values of $\nu_{12}$, which give the irreducible content in the decomposition \eqref{Decompo}, correspond to the eigenvalues of the combined Casimir operator
\begin{align*}
Q^{(12)}=[J_{0}^{(12)}]^2-J_{+}^{(12)}J_{-}^{(12)}-J_{0}^{(12)},
\end{align*}
which commutes with $J_{0}^{(12)}$, $J_{\pm}^{(12)}$, $Q^{(1)}$ and $Q^{(2)}$. Upon using \eqref{Addition-Rule}, the combined Casimir operator can be cast in the form
\begin{align}
\label{Cas-Full}
Q^{(12)}=2J_{0}^{(1)}J_{0}^{(2)}-(J_{+}^{(1)}J_{-}^{(2)}+J_{-}^{(1)}J_{+}^{(2)})+Q^{(1)}+Q^{(2)}.
\end{align}
The Clebsch-Gordan coefficients relate two possible bases for the module $V^{(\nu_1)}\otimes V^{(\nu_2)}$. On the one hand the direct product basis with vectors
\begin{align}
\label{Direct-Basis}
\ket{\nu_1,n_1}\otimes \ket{\nu_2,n_2}\equiv \ket{\nu_1,n_1;\nu_2,n_2},
\end{align}
and on the other hand, the ``coupled'' basis with vectors $\ket{\nu_{12},n_{12}}$ defined by
\begin{align}
\label{Coupled-Basis}
Q^{(12)}\ket{\nu_{12},n_{12}}=\nu_{12}(\nu_{12}-1)\ket{\nu_{12},n_{12}},\quad J_0^{(12)}\ket{\nu_{12},n_{12}}=(n_{12}+\nu_{12})\ket{\nu_{12},n_{12}}.
\end{align}
In both bases, the Casimir operators  $Q^{(1)}$, $Q^{(2)}$ act as multiples of the identity. The two bases are orthonormal and span the representation space $V^{(\nu_1)}\otimes V^{(\nu_2)}$. Hence it follows that they are related by a unitary transformation
\begin{align}
\label{CG-Decompo}
\ket{\nu_{12},n_{12}}=\sum_{n_1,\,n_2}\mathcal{C}^{\nu_1,\nu_2,\nu_{12}}_{n_1,n_2,n_{12}}\ket{\nu_1,n_1;\nu_2,n_2}.
\end{align}
By virtue of \eqref{Addition-Rule} and \eqref{Coupled-Basis}, it is clear that the condition
\begin{align*}
n_{12}+\nu_{12}=n_1+n_2+\nu_1+\nu_2,
\end{align*}
holds in the decomposition \eqref{CG-Decompo}. Since $n_{12}$ is an integer, it follows that
\begin{align}
\label{CND-2}
\nu_{12}=\nu_{1}+\nu_{2}+x,\quad n_{12}+x=n_{1}+n_{2},
\end{align}
where $x\in\{0,\ldots,N\}$ for a given value of $N=n_1+n_2$. The coefficients $\mathcal{C}^{\nu_1,\nu_2,\nu_{12}}_{n_1,n_2,n_{12}}$, which can be written
\begin{align}
\label{CG-Def}
\mathcal{C}^{\nu_1,\nu_2,\nu_{12}}_{n_1,n_2,n_{12}}=\braket{\nu_1,n_1;\nu_2,n_2}{\nu_{12},n_{12}},
\end{align}
are the Clebsch-Gordan coefficients for the positive-discrete series of irreducible representations $\mathfrak{su}(1,1)$.
\subsection{Explicit expression for the Clebsch-Gordan coefficients}
The explicit expression for the Clebsch-Gordan coefficients \eqref{CG-Def} is known \cite{Vilenkin-1991}, hence only a short derivation using a recurrence relation is presented. By definition of the coupled basis states \eqref{Coupled-Basis}, one has
\begin{align}
\label{Side-1}
\nu_{12}(\nu_{12}-1)\,\mathcal{C}^{\nu_1,\nu_2,\nu_{12}}_{n_1,n_2,n_{12}}=\Braket{\nu_1,n_1;\nu_2,n_2}{Q^{(12)}}{\nu_{12},n_{12}}.
\end{align}
On the other hand, upon using \eqref{Cas-Full} and the actions \eqref{Actions-2}, one finds
\begin{align}
\nonumber
&\Braket{\nu_1,n_1;\nu_2,n_2}{Q^{(12)}}{\nu_{12},n_{12}}=\left\{2(n_1+\nu_1)(n_2+\nu_2)\right\}\,\mathcal{C}^{\nu_1,\nu_2,\nu_{12}}_{n_1,n_2,n_{12}}
\\
\nonumber
&-\sqrt{n_1(n_1+2\nu_1-1)(n_2+1)(n_2+2\nu_2)}\,\mathcal{C}^{\nu_1,\nu_2,\nu_{12}}_{n_1-1,n_2+1,n_{12}}+\nu_1(\nu_1-1)\,\mathcal{C}^{\nu_1,\nu_2,\nu_{12}}_{n_1,n_2,n_{12}}
\\
\label{Side-2}
&-\sqrt{n_2(n_2+2\nu_2-1)(n_1+1)(n_1+2\nu_1)}\, \mathcal{C}^{\nu_1,\nu_2,\nu_{12}}_{n_1+1,n_2-1,n_{12}}+\nu_2(\nu_2-1)\,\mathcal{C}^{\nu_1,\nu_2,\nu_{12}}_{n_1,n_2,n_{12}}.
\end{align}
For a given value of $N=n_1+n_2$, taking $n_1=n$ and $n_2=N-n$, one can use the conditions \eqref{CND-2} to make explicit the dependence of $\mathcal{C}$ on $x$:
$$ 
\mathcal{C}^{\nu_1,\nu_2,\nu_{12}}_{n_1,n_2,n_{12}}=\omega\,P_{n}(x;\nu_1,\nu_2;N),
$$
where $\omega=\mathcal{C}^{\nu_1,\nu_2,\nu_1+\nu_2+x}_{0,N,N-x}$ and $P_0(x)=1$.
With these definitions, it follows from \eqref{Side-1} and \eqref{Side-2} that $P_{n}(x)$ satisfies the three-term recurrence relation
\begin{align*}
\nonumber
\lambda(x) P_{n}(x;\nu_1,\nu_2;N)&=2\{n(N-n)+\nu_2n+\nu_1(N-n)\}\,P_{n}(x;\nu_1,\nu_2;N)\\
&+W_{n}\,P_{n-1}(x;\nu_1,\nu_2;N)+W_{n+1}\,P_{n+1}(x;\nu_1,\nu_2;N).
\end{align*}
where 
$$\lambda(x)=x(x+2\nu_1+2\nu_2-1)$$
and where
\begin{align*}
W_{n}=-[n(N-n+1)(n+2\nu_1-1)(N-n+2\nu_2)]^{1/2}.
\end{align*}
Upon taking $P_{n}(x;\nu_1,\nu_2;N)=[W_{1}\ldots W_{n}]^{-1}\widehat{P}_{n}(x;\nu_1,\nu_2;N)$, one finds
\begin{align}
\label{Recu-Norm}
\lambda(x)\widehat{P}_{n}(x)=\widehat{P}_{n+1}(x)-(A_{n}+C_{n})\widehat{P}_{n}(x)+A_{n-1}C_{n}\widehat{P}_{n-1}(x),
\end{align}
where
\begin{align*}
A_{n}=(n-N)(n+2\nu_1),\quad C_{n}=n(n-2\nu_{2}-N).
\end{align*}
It is directly seen from \eqref{Recu-Norm} that the polynomials $\widehat{P}_{n}(x)$ correspond to the monic dual Hahn polynomials $R_{n}(\lambda(x);\gamma,\delta;N)$ with parameters $\gamma=2\nu_1-1$ and $\delta=2\nu_2-1$. The dual Hahn polynomials are defined by \cite{Koekoek-2010}
\begin{align}
\label{Dual-Hahn}
R_{n}(\lambda(x);\gamma,\delta;N)=\pFq{3}{2}{-n,-x,x+\gamma+\delta+1}{\gamma+1,-N}{1}.
\end{align}
Since the orthonormality condition
\begin{align*}
\sum_{\substack{\nu_{12},n_{12}\\ \nu_{12}+n_{12}=n_1+n_2+\nu_1+\nu_2}}\mathcal{C}^{\nu_1,\nu_2,\nu_{12}}_{n_1,n_2,n_{12}}\mathcal{C}^{\nu_1,\nu_2,\nu_{12}}_{n_1',n_2',n_{12}}=\delta_{n_1n_1'}\delta_{n_2n_2'},
\end{align*}
must hold, one can use the orthogonality relation of the dual Hahn polynomials to completely determine the coefficients $\mathcal{C}^{\nu_1,\nu_2,\nu_{12}}_{n_1,n_2,n_{12}}$ up to a phase factor. One finds
\begin{align}
\nonumber
\mathcal{C}^{\nu_1,\nu_2,\nu_{12}}_{n_1,n_2,n_{12}}&=\left[\frac{(2\nu_1)_{n_1}(2\nu_2)_{n_2}(2\nu_1)_{x}}{n_1!n_2!n_{12}!x!(2\nu_2)_{x}(2\nu_1+2\nu_2+2x)_{n_{12}}(2\nu_1+2\nu_2+x-1)_{x}}\right]^{1/2}\\
\label{CG-Full}
&\,\times (x+n_{12})!\; R_{n_1}(\lambda(x);2\nu_1-1,2\nu_2-1;n_1+n_2),
\end{align}
which is valid provided that the conditions \eqref{CND-2} hold. Note that one also has
\begin{align*}
\sum_{n_1,n_2}\mathcal{C}^{\nu_1,\nu_2,\nu_{12}}_{n_1,n_2,n_{12}}\mathcal{C}^{\nu_1,\nu_2,\nu_{12}'}_{n_1,n_2,n_{12}'}=\delta_{n_{12}n_{12}'}\delta_{\nu_{12}\nu_{12}'},
\end{align*}
where the sum is restricted by $n_1+n_2=n_{12}+\nu_{12}-\nu_{1}-\nu_{2}$.
\section{Overlap coefficients for the isotropic 3D harmonic oscillator}
In this section, the explicit expressions for the overlap coefficients between the Cartesian, polar and spherical bases for the states of the isotropic 3D harmonic oscillator are given. Again, these expressions are not new and can be found in \cite{Pogo-2006}. Since these results are not so readily accessible however, we rederive them here using an interpretation in terms of the Clebsch-Gordan coefficients given in \eqref{CG-Full}.
\subsection{The Cartesian/polar overlaps}
The overlap coefficients between the Cartesian $\ket{n_x,n_y,n_z}_{C}$ and polar $\ket{n_{\rho},m,n_z'}_{P}$ basis states of the oscillator are defined by
\begin{align*}
\bbraket{C}{n_x,n_y,n_z}{n_{\rho},m,n_z'}{P}.
\end{align*}
It is obvious that
\begin{align*}
\bbraket{C}{n_x,n_y,n_z}{n_{\rho},m,n_z'}{P}=\delta_{n_z,n_z'}\;\bbraket{C}{n_x,n_y,n_z}{n_{\rho},m,n_z}{P}.
\end{align*}
One has the expansion
\begin{align}
\label{Expansion-1}
\ket{n_{\rho},m,n_z}_{P}=\sum_{n_x,n_y}\bbraket{C}{n_x,n_y,n_z}{n_{\rho},m,n_z}{P}\kket{n_x,n_y,n_z}{C},
\end{align}
where the condition $n_x+n_y=2n_{\rho}+|m|$ holds since only the states in the same energy eigenspace can be related to one another. The Cartesian basis states $\ket{n_x,n_y}_{C}=\ket{n_x}\otimes \ket{n_y}$ can be identified with vectors $\ket{\nu_1,n_1;\nu_2,n_2}$ of the direct product basis for a $\mathfrak{su}(1,1)$-module $V^{(\nu_{x})}\otimes V^{(\nu_{y})}$. Indeed, it is directly checked that the operators
\begin{align}
\label{Coupled-XY}
J_0^{(x_i)}=\frac{1}{2}(a_{x_i}^{\dagger}a_{x_i}+1/2),\quad J_{+}^{(x_i)}=\frac{1}{2}(a_{x_i}^{\dagger})^{2},\quad J_{-}^{(x_i)}=\frac{1}{2}a_{x_{i}}^2,
\end{align}
with $i=1,2$, realize the $\mathfrak{su}(1,1)$ algebra and that the Cartesian states $\ket{n_{x_i}}$, with the quantum number $n_{x_i}$ either even or odd, are basis vectors for an irreducible module $V^{(\nu_{x_i})}$ with representation parameters $\nu_{x_i}=1/4$ if $n_{x_i}$ is even and $\nu_{x_i}=3/4$ if $n_{x_i}$ is odd. Hence we have the identification
\begin{align}
\label{Correspondence-1}
\ket{2\widetilde{n}_x+q_{x},2\widetilde{n}_y+q_{y}}_{C}\sim\ket{1/4+q_{x}/2, \widetilde{n}_{x};1/4+q_{y}/2, \widetilde{n}_{y}}\equiv \ket{\nu_1,n_1;\nu_2,n_2},
\end{align}
where $q_{x},q_{y}\in\{0,1\}$ and where the third quantum number $n_z$ as been suppressed from the Cartesian states in \eqref{Correspondence-1} to facilitate the correspondence with the notation used in the previous section.

The polar basis states $\ket{n_{\rho},m,n_z}_{P}$ can be identified with vectors of the ``coupled'' basis. Indeed, consider the realization of the $\mathfrak{su}(1,1)$ algebra obtained by taking
\begin{align*}
J_{0}^{(xy)}=J_{0}^{(x)}+J_{0}^{(y)},\quad J_{\pm}^{(xy)}=J_{\pm}^{(x)}+J_{\pm}^{(y)}.
\end{align*}
By definition, the states $\ket{n_{\rho},m,n_z}_{P}$ satisfy
\begin{align*}
L_{z}\ket{n_{\rho},m,n_z}_{P}=m\ket{n_{\rho},m,n_z}_{P}.
\end{align*}
Furthermore, a direct computation shows that the coupled Casimir $Q^{(xy)}$ operator can be expressed in terms of  $L_{z}$ in the following way:
\begin{align*}
Q^{(xy)}=\frac{1}{4}(L_{z}^2-1).
\end{align*}
Hence it follows that the polar basis states are eigenvectors of the combined Casimir operator $Q^{(xy)}$ with eigenvalue
\begin{align}
Q^{(xy)}\ket{n_{\rho},m}_{P}=\frac{1}{4}(m^2-1)\ket{n_{\rho},m}_{P}.
\end{align}
Since from \eqref{Expansion-1}, \eqref{Coupled-XY} and $n_x+n_y=2n_{\rho}+|m|$ one also has
\begin{align*}
J_0^{(xy)}\ket{n_{\rho},m,n_z}_{P}=\left(n_{\rho}+\frac{|m|}{2}+1/2\right)\ket{n_{\rho},m,n_z}_{P},
\end{align*}
it is seen that the polar basis states $\ket{n_{\rho},m,n_z}_{P}$ correspond to coupled $\mathfrak{su}(1,1)$ basis states of $V^{(\nu_{xy})}$ with representation parameter $\nu_{xy}=(|m|+1)/2$. One thus writes
\begin{align}
\label{Correspondence-2}
\ket{n_{\rho},m,n_z}_{P}\sim \ket{\frac{1+|m|}{2}, n_{\rho}}\equiv\ket{\nu_{12},n_{12}}.
\end{align}
The correspondence \eqref{Correspondence-1}, \eqref{Correspondence-2} can now be used to recover the overlap coefficients between the Cartesian and polar bases of the 3D isotropic harmonic oscillator. One needs to keep in mind that for $m\neq 0$, there is a sign ambiguity in \eqref{Correspondence-2} which has to be taken into account to ensure the orthonormality conditions for the overlap coefficients. One finds
\begin{align*}
\bbraket{C}{n_x,n_y,n_z}{n_{\rho},0,n_z'}{P}=e^{i\phi}\,\delta_{n_z,n_z'}\;\mathcal{C}^{\nu_1,\nu_2,\nu_{12}}_{n_1,n_2,n_{12}},
\end{align*}
for $m=0$
\begin{align*}
\bbraket{C}{n_x,n_y,n_z}{n_{\rho},m,n_z'}{P}=\frac{e^{i\phi}}{\sqrt{2}}\,\delta_{n_z,n_z'}\;\mathcal{C}^{\nu_1,\nu_2,\nu_{12}}_{n_1,n_2,n_{12}},
\end{align*}
for $m\neq 0$, where $e^{i\phi}$ is a phase factor that remains to be evaluated. The correspondence between the quantum numbers and representation parameters is given by
\begin{subequations}
\label{Identification}
\label{Iden-1}
\begin{gather}
\nu_1=1/4+q_{x}/2,\quad \nu_2=1/4+q_{y}/2,\quad \nu_{12}=(1+|m|)/2,\\
n_1=\widetilde{n_x},\quad n_2=\widetilde{n_{y}},\quad n_{12}=n_{\rho},
\end{gather}
\end{subequations}
where $n_{x_i}=2\widetilde{n}_x+q_{x}$ with $q_{x_i}\in \{0,1\}$. The remaining phase factor can be evaluated by requiring that the expansion
\begin{align*}
\Psi_{n_{\rho},m,n_z'}(\rho,\phi,z)=\sum_{n_x,n_y}\bbraket{C}{n_x,n_y,n_z}{n_{\rho},m,n_z'}{P}\;\Psi_{n_x,n_y,n_z}(x,y,z),
\end{align*}
holds for the wavefunctions. By inspection of \eqref{Cartesian-Wave} and \eqref{Polar-Wave}, one finds
\begin{align*}
e^{i\phi}=(-1)^{\widetilde{n}_x+n_{\rho}}(\sigma_{m}\,i)^{n_y},\quad\text{ with }\quad  \sigma_{m}=
\begin{cases}
1 & m\geqslant 0,\\
-1 & m<0.
\end{cases}
\end{align*}
The complete expression for the overlaps is therefore given by
\begin{align}
\bbraket{C}{n_x,n_y,n_z}{n_{\rho},m,n_z'}{P}=\delta_{n_zn_z'}\,\left(\frac{(-1)^{\widetilde{n}_x+n_{\rho}}(\sigma_{m}\,i)^{n_y}}{\sqrt{2}}\right)\,\mathcal{C}^{\nu_1,\nu_2,\nu_{12}}_{n_1,n_2,n_{12}},
\end{align}
with the identification \eqref{Iden-1} and where it is understood that the $\sqrt{2}$ factor is to be omitted when $m=0$.
\subsection{The polar/spherical overlaps}
The overlap coefficients between the polar and spherical bases are defined by
\begin{align*}
\bbraket{P}{n_{\rho},m',n_z}{n_r,\ell,m}{S}.
\end{align*}
Since both set of basis states are eigenstates of $L_{z}$, it follows that one can write
\begin{align*}
\bbraket{P}{n_{\rho},m',n_z}{n_r,\ell,m}{S}=\delta_{mm'}\, \bbraket{P}{n_{\rho},m,n_z}{n_r,\ell,m}{S}.
\end{align*}
One has the decomposition
\begin{align}
\label{Expansion-2}
\ket{n_r,\ell,m}_{S}=\sum_{n_{\rho},n_{z}}\bbraket{P}{n_{\rho},m,n_z}{n_r,\ell,m}{S}\,\ket{n_\rho,m,n_{z}}_{P},
\end{align}
where the condition $2n_{\rho}+|m|+n_{z}=2n_r+\ell$ holds since only the states with identical energies can be related to one another. The states $\ket{n_{\rho},m}$ and $\ket{n_z}$ have already been identified with basis vectors of irreducible $\mathfrak{su}(1,1)$ representations. We thus write the polar basis states $\ket{n_{\rho},m,n_z}_{P}=\ket{n_{\rho},m}\otimes \ket{n_z}$ as direct product vectors
\begin{align*}
\ket{n_{\rho},m,2\widetilde{n}_{z}+q_{z}}_{P}\sim \ket{(1+|m|)/2, n_{\rho}; 1/4+q_{z}/2, \widetilde{n}_z}\equiv\ket{\nu_1,n_1;\nu_2,n_2}.
\end{align*}
The spherical basis states $\ket{n_r,\ell,m}_{S}$ can be identified with those of the ``coupled'' basis. Indeed, consider the $\mathfrak{su}(1,1)$ algebra obtained by taking
\begin{align}
\label{SU-Coupled}
J_{0}^{((xy)z)}=J_{0}^{(xy)}+J_{0}^{(z)},\quad J_{\pm}^{((xy)z)}=J_{\pm}^{(xy)}+J_{\pm}^{(z)}.
\end{align}
By definition, the states $\ket{n_r,\ell,m}_{S}$ satisfy
\begin{align*}
\vec{L}^2\ket{n_r,\ell,m}_{S}=\ell(\ell+1)\ket{n_r,\ell,m}_{S},
\end{align*}
where $\vec{L}^2=L_{x}^2+L_{y}^2+L_{z}^2$. Furthermore, a direct computation shows that $\vec{L}^2$ and the coupled Casimir operator $Q^{((xy)z)}$ are related by
\begin{align*}
Q^{((xy)z)}=\frac{1}{4}\left(\vec{L}^2-\frac{3}{4}\right).
\end{align*}
Hence one may write
\begin{align*}
Q^{((xy)z)}\ket{n_r,\ell,m}_{S}=(\ell/2+3/4)(\ell/2-1/4)\ket{n_r,\ell,m}_{S}.
\end{align*}
Since from \eqref{Expansion-2}, \eqref{SU-Coupled} and the condition $2n_r+\ell=2n_{\rho}+|m|+n_z$ one has
$$
J_{0}^{((xy)z)}\ket{n_r,\ell,m}_{S}=\{n_r+(\ell+3)/2\}\ket{n_r,\ell,m}_{S},
$$ 
it follows that the states of the spherical basis correspond to coupled $\mathfrak{su}(1,1)$ states 
\begin{align*}
\ket{n_r,\ell,m}_{S}\sim \ket{\frac{\ell+3/2}{2},n_r}\sim \ket{\nu_{12},n_{12}}.
\end{align*}
Using this identification, one writes
\begin{align*}
\bbraket{P}{n_{\rho},m,n_z}{n_r,\ell,m}{S}=e^{i\psi}\delta_{mm'}\,\mathcal{C}^{\nu_1,\nu_2,\nu_{12}}_{n_1,n_2,n_{12}},
\end{align*}
where
\begin{subequations}
\label{Iden-2}
\begin{gather}
\nu_1=\frac{1+|m|}{2},\quad \nu_2=\frac{1/2+q_{z}}{2},\quad \nu_{12}=\frac{\ell+3/2}{2},\\
n_1=n_{\rho},\quad n_2=\widetilde{n}_{z},\quad n_{12}=n_{r},
\end{gather}
\end{subequations}
and with $n_{z}=2\widetilde{n}_z+q_{z}$. The phase factor $e^{i\psi}$ can be determined by requiring that the expansion formula
\begin{align*}
\Psi_{n_r,\ell,m}(\rho,\theta,\phi)=\sum_{n_{\rho},n_z}\bbraket{P}{n_{\rho},m,n_z}{n_r,\ell,m}{S}\; \Psi_{n_{\rho},m,n_z}(\rho,\phi,z),
\end{align*}
holds for the wavefunctions. Upon inspecting \eqref{Polar-Wave} and \eqref{Spherical-Wave}, one finds that $e^{i\psi}=i^{m+|m|}$. Hence the following expression holds
\begin{align}
\bbraket{P}{n_{\rho},m,n_z}{n_r,\ell,m'}{S}=\delta_{mm'}\,i^{m+|m|}\,\mathcal{C}^{\nu_1,\nu_2,\nu_{12}}_{n_1,n_2,n_{12}},
\end{align}
with the identification \eqref{Iden-2}. Note that one has also
\begin{align*}
\bbraket{S}{n_r,\ell,m'}{n_{\rho},m,n_z}{P}=\delta_{mm'}\,(-i)^{m+|m|}\,\mathcal{C}^{\nu_1,\nu_2,\nu_{12}}_{n_1,n_2,n_{12}},
\end{align*}
since the Clebsch-Gordan coefficients $\mathcal{C}^{\nu_1,\nu_2,\nu_{12}}_{n_1,n_2,n_{12}}$ are real.
\section{Conclusion}
To sum up, we have obtained a new explicit formula for the bivariate Krawtchouk polynomials in terms of the standard (univariate) Krawtchouk and dual Hahn polynomials. Furthermore, the explicit expressions for the overlap coefficients of the isotropic oscillator have been rederived using a correspondence with the Clebsch-Gordan problem of $\mathfrak{su}(1,1)$. 

In \cite{Genest-2013-1}, the results obtained using $SO(3)$ were seen to extend directly to higher dimensions and indeed the $d$-variable Krawtchouk polynomials can be interpreted as matrix elements of unitary reducible $SO(d+1)$ representations on (Cartesian) oscillator states. The main result \eqref{Main-Formula} obtained here for the bivariate Krawtchouk polynomials can also be generalized to $d$ variables. The derivation is similar in spirit to the one presented here but is quite technical. We now outline how this generalization proceeds. 

Let $R\in SO(d+1)$. The matrix elements of the reducible $SO(d+1)$ unitary representation \eqref{Unirep} in the Cartesian basis of the $\mathcal{E}=N+d/2$ energy eigenspace of the $(d+1)$-dimensional isotropic harmonic oscillator are expressed as follows \cite{Genest-2013-1}:
\begin{align*}
\Braket{i_1,\ldots,i_{d+1}}{U(R)}{n_1,\ldots,n_{d+1}}=W_{i_1,\ldots,i_{d};N}\,P_{n_1,\ldots,n_{d}}(i_1,\ldots,i_{d};N),
\end{align*}
where $P_{n_1,\ldots,n_{d}}(i_1,\ldots,i_{d};N)$ are the multivariate Krawtchouk polynomials and where 
$$
W_{i_1,\ldots,i_{d};N}=\binom{N}{i_1,\ldots,i_{d}}^{1/2}R_{1,d+1}^{i_1}\ldots R_{d,d+1}^{i_{d}}R_{d+1,d+1}^{N-i_1-\ldots -i_{d}},
$$
with $\sum_{k=1}^{d+1}i_{k}=\sum_{k=1}^{d+1}n_{k}=N$. The decomposition of this $SO(d+1)$ representation in irreducible components can be accomplished by a passage to a canonical basis which corresponds to the separation of variables of the Schr\"odinger equation in hyperspherical coordinates \cite{Avery-2010}. These basis states are denoted by $\ket{n_r,\lambda,\mu_1,\ldots \mu_{d-1}}$ with $n_r,\in \mathbb{N}$ and $\lambda\geqslant \mu_1\geqslant\cdots \geqslant |\mu_{d-1}|\geqslant 0$. They are eigenstates of the $(d+1)$-dimensional harmonic oscillator Hamiltonian with energy $\mathcal{E}=2n_r+\lambda+d/2$ and the corresponding wavefunctions can be expressed in terms of Laguerre polynomials and hyperspherical harmonics \cite{Avery-1994,Wen-1985}. These states form a basis for (class 1) irreducible representations of $SO(d+1)$ \cite{Vilenkin-1991}. They are eigenvectors of the quadratic Casimir operator of $SO(d+1)$ with eigenvalue $\lambda(\lambda+d-1)$ and of the quadratic Casimir operators of each element in the canonical subgroup chain $SO(d+1)\supset SO(d)\supset\cdots \supset SO(2)$ with eigenvalues $\mu_1(\mu_1+d-2),\mu_2(\mu_2+d-3)\ldots,\mu_{d-1}^2$ \cite{Avery-2010}. This is the origin of the parameters $\mu_i$, $i=1,\ldots,d-1$. For a given $N$, the $SO(d+1)$ representation on the eigenstates of the $(d+1)$-dimensional oscillator contains once, each and every (class one) irreducible representation of $SO(d+1)$ with $\lambda=N,N-2,\ldots,0,1$ depending on the parity. This decomposition is equivalent to the decomposition of the quasi-regular representation of $SO(d+1)$ \cite{Vilenkin-1991}. Upon introducing the states corresponding to separation in hyperspherical coordinates, one is led to the decomposition formula
\begin{align*}
P_{n_1,\ldots,n_{d}}(i_1,\ldots,i_{d};N)&=W_{i_1,\ldots,i_{d};N}^{-1}\sum_{n_r,\lambda}\sum_{\mathbf{\mu},\mathbf{\mu'}}\Braket{n_r,\lambda,\mu_1',\ldots,\mu_{d-1}'}{U(R)}{n_r,\lambda,\mu_1,\ldots,\mu_{d-1}}
\\
&\quad \braket{i_1,\ldots,i_{d+1}}{n_r,\lambda,\mu_1',\ldots,\mu_{d-1}'}\braket{n_r,\lambda,\mu_1,\ldots,\mu_{d-1}}{n_1,\ldots,n_{d+1}},
\end{align*}
where $2n_r+\lambda=N=i_1+\cdots+i_{d+1}=n_1+\cdots+n_{d+1}$ and where $\mu$ denotes the multi-index $(\mu_1,\ldots,\mu_{d-1})$ with $\lambda\geqslant \mu_1\geqslant \mu_2 \geqslant\cdots \geqslant |\mu_{d-1}|\geqslant 0$. The overlap coefficients can be evaluated as sums of products of $d$ $\mathfrak{su}(1,1)$ Clebsch-Gordan coefficients using successive recouplings of the quantum numbers, as was done in Section 3. The matrix elements  $\Braket{n_r,\lambda,\mu_1',\ldots,\mu_{d-1}'}{U(R)}{n_r,\lambda,\mu_1,\ldots,\mu_{d-1}}$ are very involved. They can be evaluated only recursively using the canonical subgroup chain of $SO(d+1)$. See \cite{Vilenkin-1991} for details.

\section*{Acknowledgements}
The authors thank  G.S. Pogosyan for exchanges on interbasis expansion coefficients. The research of L.V. is supported by the Natural Sciences and Engineering Council of Canada (NSERC). V.X.G. holds an Alexander-Graham-Bell fellowship from NSERC.

\section*{References}
\footnotesize

\end{document}